\documentclass[twocolumn,amsmath,amssymb]{revtex4} 
\usepackage{graphicx}
\usepackage{dcolumn}
\usepackage{bm}
\begin{document} 
 
\title{Graphene, Nobel Prize and All that Jazz 
\footnote{To be published in {\it Physics in Canada}}} 
\author{Tapash Chakraborty$^\ddag$} 
\affiliation{Department of Physics and Astronomy, 
University of Manitoba, Winnipeg, Canada R3T 2N2} 
 
\date{\today} 
\begin{abstract} 
\end{abstract} 
\maketitle 
 
\begin{figure} 
\begin{center}\includegraphics[width=6cm]{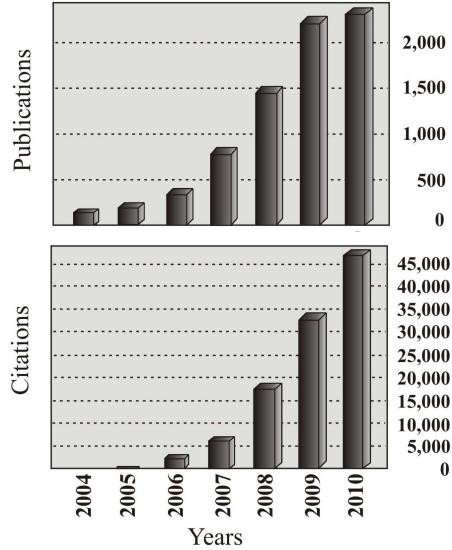} 
\end{center} 
\caption{The impact of graphene in the scientific community 
can be gauged from the number of publications and citations 
on graphene per year and the number of citations (based on 
the database of the Web of Science). 
} 
\label{Fig_pub} 
\end{figure}

Graphene, a single atomic layer of graphite, first isolated in 2004 
\cite{novo_science}, has made a quantum leap in the exploration of 
the physics of two-dimensional electron systems \cite{review,my_pic}. 
Since the initial report of its discovery, many thousands of papers 
have been published (Fig.~\ref{Fig_pub}), attempting to explain every 
aspect of the exotic electronic properties of this system. The 
graphene euphoria has culminated with the 2010 Nobel Prize in physics 
being awarded jointly to Andre Geim and Konstantin Novoselov of 
the University of Manchester, UK, ``for groundbreaking experiments 
regarding the two-dimensional material graphene". But, what are 
the properties of graphene, and how was it made? Why it is so exciting 
for so many researchers, and why the Nobel Prize? 
 
\begin{figure} 
\begin{center}\includegraphics[width=6cm]{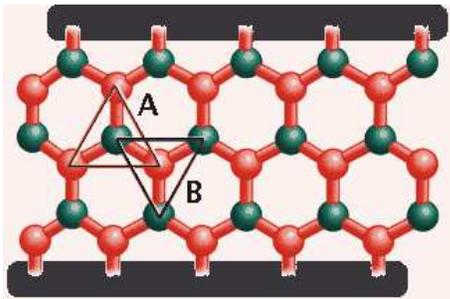} 
\end{center} 
\caption{Graphene lattice. The unit cell contains two atoms atoms 
A and B. 
} 
\label{Fig_lattice} 
\end{figure}

The saga of the exotic electronic properties of graphene actually 
began in Canada, when the band structure of graphene was first 
reported in a seminal paper by Philip R. Wallace (then at the 
National Research Council of Canada, Chalk River) \cite{wallace}. 
Graphene is a single sheet of carbon atoms arranged in a honeycomb 
(hexagonal) lattice (a molecular chicken wire where one carbon 
atom sits at each 120$^\circ$ corner) (Fig.~\ref{Fig_lattice}). 
This material is perhaps the ultimate two-dimensional system 
possible, with very unique electronic properties that are entirely 
different (and unexpected) from those of conventional two-dimensional 
systems \cite{ando}. Graphene is a bipartite lattice made up of 
two interpenetrating triangular sublattices. There are two carbon 
atoms (commonly referred to as A and B) per unit cell. Each carbon 
atom has one $s$ orbital and two in-plane $p$ orbitals which make 
up the strong covalent bonds responsible for the mechanical stability 
of the graphene sheet. The remaining $p^{}_z$ orbital, pointing out 
of the graphene sheet, form the conduction and valence bands 
with the neighboring $p^{}_z$ orbitals. 
 
\begin{figure} 
\begin{center}\includegraphics[width=7cm]{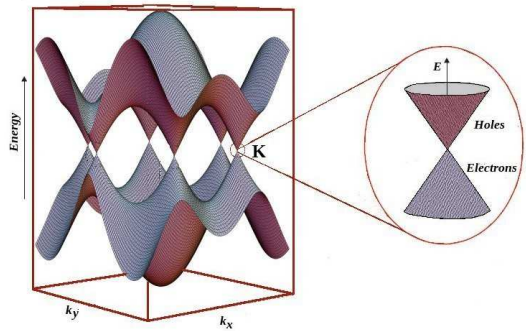} 
\end{center} 
\caption{Energy dispersion relation of graphene. The lower band is 
completely filled and meets the totally empty band at the K point. 
} 
\label{Fig_band} 
\end{figure}

\begin{figure} 
\begin{center}\includegraphics[width=7cm]{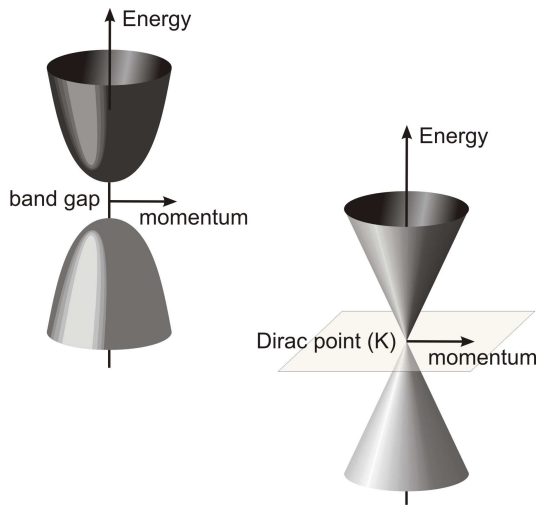} 
\end{center} 
\caption{Comparison of the energy dispersion of electrons in 
conventional two-dimensional systems (left) and Dirac cones 
(right). 
} 
\label{Fig_cone} 
\end{figure}

The dynamics of electrons in graphene can be described by a 
tight-binding Hamiltonian with nearest-neighbor hopping. Using 
this approach, Wallace derived the band structure of graphene 
as shown in Fig.~\ref{Fig_band}. There are six points where 
the valence (upper) and conduction (lower) bands meet. Two of 
those points are inequivalent and are usually denoted by K and 
K$^\prime$ \cite{review}. In pristine graphene, the Fermi level 
lies right at the meeting points, rendering graphene a zero-band 
gap semiconductor. For reasons that will be clear below, these 
points are known as Dirac (or charge neutrality) points. Close 
to the Dirac points the charge carrier dispersion relation is, 
${\cal E}\approx\hbar v^{}_{\rm F} \vert{\bf k}\vert$, where the 
momentum $\bf k$ is measured with respect to the K (or K$^\prime$) 
point. Here, $v^{}_{\rm F}\approx c/300$, $c$ being the speed 
of light, is the Fermi velocity. Clearly, the charge carriers 
behave as if they were massless relativistic particles (obeying the 
Dirac equation rather than the Schr\"odinger equation), and display 
properties that are completely different from those in conventional 
metals or semiconductors, where the energy dispersion is parabolic 
(Fig.~\ref{Fig_cone}). 
 
\begin{figure} 
\begin{center}\includegraphics[width=7cm]{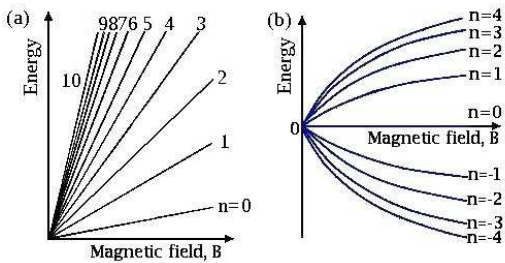} 
\end{center} 
\caption{The Landau levels for (a) electrons in conventional 
two-dimensional systems and (b) in graphene. 
} 
\label{Fig_landau} 
\end{figure}

In the presence of an external magnetic field, the energies 
of electrons in conventional two-dimensional systems are 
quantized as Landau levels that are linearly 
proportional to the magnetic field (Fig.~\ref{Fig_landau}). 
In the quantum limit, i.e., at low temperatures when the 
thermal energy is much less than the level spacing, and at high 
magnetic fields when the Landau level spacings are large, the 
integer quantum Hall effect (QHE) was discovered in 1980 
\cite{book} (Nobel prize in 1985). In the extreme quantum limit, 
i.e., at very high fields and very low temperatures, and in 
samples with very high electron mobility, the fractional QHE 
was discovered in 1982 \cite{book} (Nobel prize in 1998). How 
does the situation differ in the case of graphene? Back in 
1956, McClure reported \cite{mcclure} that the linear dispersion 
of energy dictates a very different type of Landau level: ones 
with square-root dependence on the magnetic field, 
and presence of a level at zero energy that is independent of 
the magnetic field (Fig.~\ref{Fig_landau}). In this case, an 
unusual QHE, the so-called half-integer QHE was expected 
\cite{sharapov}. Experimental observation of this half-integer 
QHE \cite{novo_nature,kim_nature} firmly established the 
existence of massless Dirac type of behavior of charge carriers 
in graphene. This momentous observation sparked intense 
interest in the electronic properties of graphene that we 
are still witnessing today. 
 
Graphite is composed of stacked layers of graphene with weak 
interlayer coupling by the van der Waals force. Isolated sheets 
of graphene which eluded us for many years, were actually 
prepared by a remarkably simple, and yet efficient method by 
Geim and Novoselov \cite{novo_science}. The process is called 
mechanical exfoliation (a formal name for a pedestrian `scotch 
tape' technique), where graphite flakes are produced by repeated 
peeling of graphite with adhesive tapes until some of the flakes 
are found to be monolayers (as determined by optical microscopy). 
The flakes are then deposited on a silicon wafer with an SiO$_2$ top 
layer of thickness 300 nm. The one-atom-thick bits of graphene 
were found to remain surprisingly stable even at room temperature 
\cite{novo_science}. 
 
The Scotch-tape "synthesis" method, prosaic in the face of 
epitaxial technology, isn't likely to produce large quantities 
of graphene. In the race to achieve graphene in large quantities 
various other techniques have been reported in the literature 
\cite{synthesis}. The goal to grow uniform, large piece of 
graphene that are suitable for graphene-based electronics, has 
achieved only a limited success. 
 
%
 
Extensive research on graphene in the past five years have accumulated 
a wealth of information on the properties of graphene. It (a) shows 
ballistic transport over sub-micron scales with very high mobilities 
when suspended \cite{morozov,communication}, (b) can withstand large 
current densities \cite{novo_science}, (c) is almost impermeable to 
gases \cite{bunch}, (d) is chemically stable, and (e) has high thermal 
conductivity, among others. These appealing traits make graphene 
a very promising candidate for nano-electronics applications. 
 
Can graphene replace silicon based transistors? The carrier mobilities in 
graphene are hundreds of times higher than in silicon chips used today. 
However, there is a `slight' problem with graphene, it has no band 
gap! A gap between the conduction and valence band allows semiconductors to 
be easily turned on and off. On the other hand, the absence of a band 
gap allows absorption of light in a large range of the electromagnetic 
spectrum. Therefore it opens a huge potential for applications of graphene 
in electronic-photonic devices. The problems related to the absence of 
a band gap in graphene, in its use in field-effect transistors, can be 
circumvented by using graphene nanoribbons instead \cite{han,avouris}. 
A thin film of graphene has very high transparency and combined with 
high electrical conductivity makes graphene an ideal candidate for 
transparent conductive coatings. This has already been demonstrated 
in graphene based solar cells \cite{solar}. Graphene membranes, just 
one-atom thick were shown to stretch like a balloon, yet strong enough 
to contain gases under several atmosphere of pressure without 
bursting \cite{bunch}. Chemical modification of graphene is a promising 
route to create a band gap. Adsorption of hydrogen on a graphene surface 
creates a new system, named graphane \cite{graphane}. The electronic 
properties of the new material change markedly by having a band gap. 
The new system might be the testing ground for unique magnetic 
properties \cite{julia} suitable for spin based electronics. 
 
Quantum dots (QDs), or {\it artificial atoms} \cite{QD_book} are 
one of the most intensely studied systems in condensed matter physics, 
where the fundamental effects related to various quantum phenomena in 
confined geometries can be studied but with the unique advantage that 
the nature of the confinement and the electron density can be tuned 
externally. However, much of the interest in this system derives 
from its enormous potential for applications, ranging from novel lasers 
to quantum information processing. While the majority of the QD systems 
investigated are based on the semiconductor heterostructures, research 
on QDs created from graphene have been reported 
\cite{review,QD_graphene,qdots_expt} recently. 
 
By stacking one more layer on graphene, one gets bilayer 
graphene, which is a fascinating and complex system in its 
own right, distinct from both the monolayer and traditional 
two-dimensional electron systems, even though it shares some 
characteristics of each \cite{review}. It has a quadratic 
low-energy band structure and the charge carriers are massive, 
unlike the massless nature in the case of monolayers \cite{bilayer}. 
One notable feature of bilayer graphene is the ability to open 
a tunable band gap by engineering a potential difference 
between the two layers, suitable for the construction of devices. 
Understanding the role electron-electron interactions play in 
bilayer graphene might be the key to explain many of the 
important features observed in this system \cite{david}. 
 
As stated above, the observation of a half-integer QHE in 
graphene opened the floodgate of enthusiasm for graphene-related 
research. Interestingly, the ``other" QHE, the fractionally 
quantized Hall effect was explored by this author theoretically 
for monolayer \cite{vadim_fqhe} and bilayer \cite{vadim_bilayer} 
graphene. The effect has been recently observed experimentally 
in monolayer graphene \cite{fqhe_kim}. This is an important 
milestone in graphene research, in particular, for 
investigation of highly correlated electrons in graphene. 
 
The award of this year's Nobel prize is a strong testament to 
the fact that in these days when ground breaking discoveries 
in science are thought to come exclusively from multi-billion 
dollar particle accelerators searching for the elusive bosons, 
or telescopes in space zooming in on truly unreachable exoplanets, 
fundamental and profound discoveries can also be made with a 
tiny piece of graphite and ordinary sticky tape. Interestingly, 
this award has also been given for a reason that is totally 
opposite to not-so-unrelated discoveries of the integer and 
fractional quantum Hall effects. Those discoveries were recognized 
by the awards which somewhat overshadowed the extraordinary 
developments in materials where those effects were discovered 
and how those materials advanced the semiconductor technology. 
In graphene, on the other hand, thanks to pioneers such as 
Wallace and McClure, there were plenty of valuable insights 
already available. The challenge was to create materials where 
those predictions could be verified. Geim and Novoselov 
were successful in doing just that. Award of this Nobel 
Prize highlights the importance of graphene in nanotechnology. 
We will have to wait and see if graphene the `wunder material' 
would actually face up to the mighty challenger, silicon, and 
surpass its prowess in future nanoscale electronic devices. 
 
My research work cited in this article was supported by the 
Canada Research Chairs Program. I wish to thank Dr. Julia 
Berashevich for her help in preparing this article. I also 
thank Vadim Apalkov for a critical reading of the article. 
 

\end{document}